\documentclass[12pt,a4paper]{article}
\usepackage[utf8]{inputenc}
\usepackage{amsmath}
\usepackage{amsfonts}
\usepackage{amssymb}
\usepackage{makeidx}
\usepackage{graphicx}

\begin{document}

\begin{center}
{\Large \bf Tunneling time in space fractional quantum mechanics
  }

\vspace{1.3cm}

{\sf   Mohammad Hasan  \footnote{e-mail address: \ \ mhasan@isac.gov.in, \ \ mohammadhasan786@gmail.com}$^{,3}$,
 Bhabani Prasad Mandal \footnote{e-mail address:
\ \ bhabani.mandal@gmail.com, \ \ bhabani@bhu.ac.in  }}

\bigskip

{\em $^{1}$ISRO Satellite Centre (ISAC),
Bangalore-560017, INDIA \\
$^{2,3}$Department of Physics,
Banaras Hindu University,
Varanasi-221005, INDIA. \\ }

\bigskip
\bigskip

\noindent {\bf Abstract}

\end{center}
We calculate the time taken by a wave packet to travel through a classically forbidden region of space in space fractional quantum mechanics. We obtain the  close form expression of tunneling time from a rectangular barrier by stationary phase  method. We show that tunneling time depends upon the width $b$ of the barrier for $b \to \infty$ and therefore Hartman effect doesn't exist in space fractional quantum mechanics. Interestingly we found  that the tunneling time monotonically reduces with increasing $b$. The tunneling time is smaller in space fractional quantum mechanics as compared to the case of standard quantum mechanics. We recover the Hartman effect of standard quantum mechanics as a special case of space fractional quantum mechanics.

\medskip
\vspace{1in}
\newpage

\section{Introduction}
The concept of fractals  in quantum mechanics was introduced in the context of path integral (PI) formulation of quantum mechanics
\cite{feynman}. In the PI formulation, the path integrals are taken over Brownian paths which results in Schrodinger equation of motion. Nick Laskin generalized the path integral approach of quantum mechanics by considering the path integrals over Levy flight paths \cite{1,2}. The Levy flight paths are characterized by a parameter $ \alpha $ known as Levy index . For $\alpha=2$ the Levy flight paths are Brownian paths. This natural generalization of quantum mechanics is termed as space fractional quantum mechanics (SFQM) and is governed by fractional Schrodinger equation (or Laskin equation). The range of $\alpha$ in SFQM is limited to $1<\alpha \leq2$ \cite{1}. All the results of  standard quantum mechanics are the special case  of SFQM with Levy index $\alpha=2$.  The field of SFQM has grown fast over the last one and a half decade and various applications of fractional generalization of quantum mechanics have been discussed \cite{3,4,5,6,longhi_fractional_optics}. 
\paragraph{}
The tunneling of a particle to cross the classically forbidden region of space is one of the most fundamental problem of quantum mechanics. The propagation of evanescent wave through a potential barrier has long been studied \cite{condon, eisenbund_dissertation, wigner_1955, david_bohm_1951}. However calculating the duration  of time that a particle spends within the spatial extent of classically forbidden potential during tunneling is  remained an open problem till date. One of the most interesting and puzzling features of tunneling time studies is the saturation of tunneling time with the thickness of classically forbidden region of space. This phenomena is known as Hartman effect. In $1962$, Hartman studied the tunneling time by stationary phase method for metal-insulator-metal sandwich and showed that for a thick barrier the tunneling time is independent of the thickness \cite{hartman_paper}. Later Fletcher independently showed that an evanescent wave travelling through a thick barrier shows saturation of tunneling time with the thickness \cite{fletcher}. These exciting theoretical results prompted a series of important experiments to test the validity of Hartman effect. The experiments were performed with microwave \cite {nimtz, ph}and optical ranges \cite {sattari, longhi1} for single, double and multiple gratings . It was noticed that the tunneling time in all such experiments doesn't show any change with the width of the tunneling region.  The tunneling time was also experimentally studied with double barrier optical gratings \cite{longhi1}. It was found that the tunneling time is paradoxially short and also independent of the gap between the two optical gratings. This result was also in favour of the generalized Hartman effect in which the tunneling time is also independent of the distance between two potential barriers \cite{generalized_hartman} and the interbarrier separation for the case of multiple barrier \cite{esposito_multi_barrier}. The problem of tunneling time was also studied for complex potentials. For complex potentials associated with elastic and inelastic channels, the tunneling time was found to saturate with the thickness of the barrier for the case of weak absorptions \cite{dutta, our}.
\paragraph{}  
The tunneling problem of space fractional quantum mechanics has been solved for various potential configurations by many authors. E. C. de Oliveira et al \cite{8} provided the reflection and transmission coefficient for delta and double delta potentials . They found the existence of zero energy transmission from such a system. For a rectangular barrier the expression of reflection and transmission coefficients of a particle are calculated by Guo et al. \cite{4}. The formulation developed by Griffith et al \cite{griffith} for the scattering coefficients of a periodic system are used in \cite{tare_2014} to study the transmission from Dirac comb and periodic rectangular barrier in space fractional quantum mechanics.
\paragraph{}
Considering the growing interest and development of SFQM as well as in the tunneling time, we explore the tunneling time by stationary phase method in the domain of SFQM for a classically forbidden rectangular barrier. We obtain the close form expression for tunneling time to show that with an increase in barrier thickness tunneling time does depend upon the width of the barrier for $1 < \alpha < 2$. This shows that Hartman effect doesn't exist in space fractional quantum mechanics. For $\alpha=2$ we recover the Hartman effect of standard quantum mechanics. For $1 < \alpha < 2$, the tunneling time is smaller than the case of $\alpha=2$. We provide the explanation for this result. However it is found that in space fractional quantum mechanics, the tunneling time first attain a maxima as the thickness of the barrier increases and then monotonically decreases. The monotonic decrease in tunneling time of a wave packet after a certain barrier thickness is even more paradoxical result than the Hartman effect.   
\paragraph{}
This paper is organised as follows: in  section \ref{intro_tt} we outline  the stationary phase  method of calculating the tunneling time and the Hartman effect of standard quantum mechanics. We discuss the fractional Schrodinger equation in \ref{fse}. In section \ref{tt_in_sfqm} we calculate the tunneling time in SFQM and show that Hartman effect doesn't exist in the domain of SFQM. Finally the results are discussed in section \ref{results_discussions}.
\section{Tunneling time in standard quantum mechanics}
\label{intro_tt}
In this section we present the stationary phase  methodology to calculate the time taken by a free particle to travel the classically forbidden region of space \cite{dutta_roy_book}. The stationary phase  method defines the tunneling time as the time difference between the peak of the incoming and outgoing  localized wave packet as it traverses the potential barrier. To find the tunneling time $\tau$ consider the time evolution of a localized wave packet $G_{k_{0}} (k)$ given by normalized Gaussian function having peak at mean momentum $\hbar k_{0}$
\begin{equation}
\int G_{k_{0}} (k)e^{i(kx-\frac{Et}{\hbar})}dk
\label{localized_wave_packet}
\end{equation}
where wave number $k=\sqrt{2mE}$. The wave packet propagates towards positive $x$ direction. Due to interaction of wave packet with the barrier, the transmitted wave packet would be
\begin{equation}
\int G_{k_{0}} (k) \vert A(k) \vert e^{i(kx-\frac{Et}{\hbar} +\Phi(k))}dk
\label{emerged_wave_packet}
\end{equation}
where $A(k)=\vert A(k) \vert e^{i\Phi (k)}$ is the transmission coefficient through the rectangular  potential barrier $V(x)$ ($V(x)=V $ for $0 \leq x \leq b$ and zero elsewhere). According to stationary phase method the tunneling time $\tau$ is given by
\begin{equation}
\frac{d}{dk} \left( kb-\frac{E\tau}{\hbar} +\Phi(k) \right)=0
\label{spm_condition}
\end{equation}
This gives the tunneling time expression as
\begin{equation}
\tau= \hbar \frac{d \Phi(E)}{dE} +\frac{b}{(\frac{\hbar k}{m})}
\label{phase_delay_time}
\end{equation}
For a square barrier potential $V(x)=V$ confined over the region $0 \leq x \leq b$ and zero elsewhere, the tunneling time is
\begin{equation}
\tau= \hbar \frac{d}{dE} \tan^{-1} \left( \frac{k^{2}-q^{2}}{2kq} \tanh{qb}\right)
\end{equation}
In the above equation, $q= \sqrt{2m(V-E)}/\hbar$. We observe that $\tau \rightarrow 0$ as $b \rightarrow 0$ as expected, however when $b \rightarrow \infty$, $\tau=\frac{2m}{\hbar qk}$, i.e. tunneling time is independent of the width of the barrier $b$ for a sufficiently opaque barrier. Hence in the units  $2m=1$, $\hbar=1$, $c=1$
\begin{equation}
\lim_{b\rightarrow \infty} \tau \sim \frac{1}{qk}
\label{tt_qm}
\end{equation}
This is the famous Hartman effect, i.e the tunneling time is independent of the width of the barrier for sufficiently thick barrier.
\section{The fractional Schrodinger equation}
\label{fse}
The fractional Schrodinger equation in one dimension is
\begin{equation}
i \hbar \frac{\partial \psi (x,t)}{\partial x}= H_{\alpha} (x,t) \psi(x,t) \ \ \ \ , 1<\alpha \leq2
\label{tdfse}
\end{equation}
Where $H_{\alpha} (x,t)$ is the fractional Hamiltonian operator and is expressed through Riesz fractional derivative $(-\hbar^{2} \Delta)^{\alpha/2}$ as
\begin{equation}
H_{\alpha} (x,t)=D_{\alpha} (-\hbar^{2} \Delta)^{\frac{\alpha}{2}} +V(x,t)
\end{equation}
$D_{\alpha}$ is a constant and $\Delta=\frac{\partial^{2}}{\partial x^{2}}$. The Riesz fractional derivative of the wave function $\psi(x,t)$ is defined through its Fourier transform $\tilde{\psi}(p,t)$ as
\begin{equation}
(-\hbar^{2}\Delta)^{\frac{\alpha}{2}} \psi(x,t)=\frac{1}{2\pi \hbar} \int_{-\infty} ^{\infty} { \tilde{\psi}(p,t)\vert p \vert ^{\alpha} e^{\frac{ipx}{\hbar}}dp } 
\label{Riesz_fractional_derivative}
\end{equation}
The Fourier transform of $\psi(x,t)$ is
\begin{equation}
\tilde{\psi}(p,t)= \int_{-\infty} ^{\infty} \psi(x,t) e^{-i\frac{px}{\hbar}} dx
\end{equation}
and
\begin{equation}
\psi(x,t)=\frac{1}{2\pi \hbar} \int_{-\infty} ^{\infty} \tilde{\psi}(p,t) e^{i\frac{px}{\hbar}} dp
\end{equation}
When potential $V(x,t)$ is independent of time we have the time independent fractional Hamiltonian operator $H_{\alpha}(x)=D_{\alpha} (-\hbar^{2} \Delta)^{\frac{\alpha}{2}} +V(x)$. In this case the time independent fractional Schrodinger equation is
\begin{equation}
D_{\alpha} (-\hbar^{2} \Delta)^{\frac{\alpha}{2}}\psi(x)+V(x)\psi(x)=E\psi(x)
\label{tifse}
\end{equation} 
where $E$ is the energy of the particle and $\psi(x,t)=\psi(x)e^{-\frac{iEt}{\hbar}}$.
\section{Tunneling time in space fractional quantum mechanics}
\label{tt_in_sfqm}
\begin{figure}
\begin{center}
\includegraphics[scale=0.6]{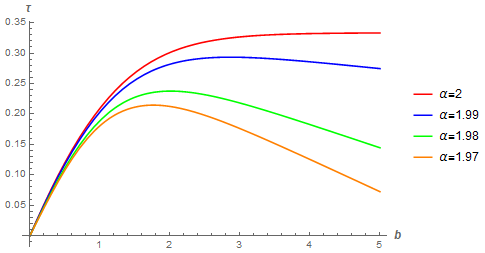}
\caption{Variation of tunneling time $\tau$ with width `$b$' of the barrier for different values of $\alpha$. It is seen that for $\alpha<2$ , $\tau$ first increase with $b$ then begins to decrease with increasing $b$. For $\alpha=2$ we recover the well known Hartman effect. In the plot $V=10$, $E=9$ and $u=10^{-4}c$, $c=1$.} 
\label{fig_alpha}
\end{center}
\end{figure}

\begin{figure}
\begin{center}
\includegraphics[scale=0.7]{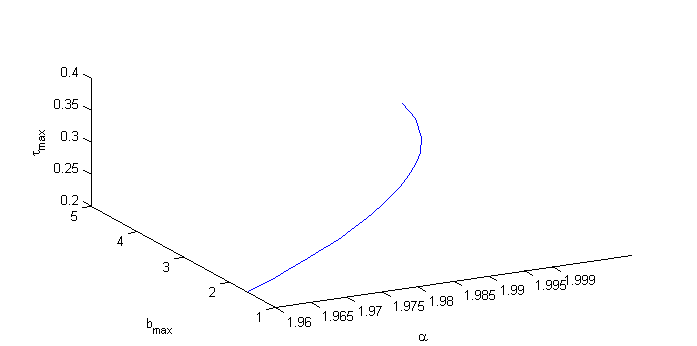}
\caption{Variation of peak tunneling time $\tau_{max}$ with $\alpha$ along with $b_{max}$. $b_{max}$ is the value of $b$ at which $\tau=\tau_{max}$.  The parameters are the same as of Fig \ref{fig_alpha}. It is seen that the particle tunnel faster with decreasing $\alpha$.}
\label{fig_peak}
\end{center}
\end{figure}

For a square barrier potential $V(x)=V$ confined over the region $0 \leq x \leq b$ and zero elsewhere, the space-fractional Schrodinger equation is
\begin{equation}
D_{\alpha} (-\hbar^{2} \Delta)^{\frac{\alpha}{2}}\psi(x)+V\psi(x)=E\psi(x)
\label{square_barrier_laskin}
\end{equation}  
The general solution of Eq. \ref{square_barrier_laskin} has the following form
\begin{equation}
\psi(x)=
	\begin{cases}
	Ae^{ik_{\alpha }x}+Be^{-ik_{\alpha }x},	         & \text{$x<0$}\\
	C \cos{\overline{k}_{\alpha}x}+D \sin{\overline{k}_{\alpha}x},      & \text{$x<0<b$}\\
	Fe^{ik_{\alpha }x}+Ge^{-ik_{\alpha }x},      & \text{$x>b$}\\
	\end{cases}
\end{equation} 
where,
\begin{equation}
k_{\alpha}=\left( \frac{E}{D_{\alpha} \hbar^{\alpha}}\right)^{\frac{1}{\alpha}}
\end{equation}
and
\begin{equation}
\overline{k}_{\alpha}=\left( \frac{E-V}{D_{\alpha} \hbar^{\alpha}}\right)^{\frac{1}{\alpha}}
\end{equation}
The transmission coefficient for a particle from a rectangular barrier in SFQM has been found in Ref \cite{4} and the transfer matrix for the barrier is given in the Ref. \cite{tare_2014}. For our problem we find it more suitable to write the transmission coefficient as 
\begin{equation}
\overline{t}=\frac{e^{-ik_{\alpha}b}}{{\cos{\overline{k}_{\alpha}b}-i\mu \sin{\overline{k}_{\alpha}b}}}
\label{tl_sSFQM}
\end{equation}
where,
\begin{equation}
\mu=\frac{1}{2}\left(\varepsilon+\frac{1}{\varepsilon} \right)
\label{mu_formula}
\end{equation}
\begin{equation}
\varepsilon=\left( \frac{k_{\alpha}}{\overline{k}_{\alpha}}\right)^{\alpha-1}
\label{epsilon_formula}
\end{equation}
For classically forbidden case, $E<V$, the expression for tunneling coefficient is
\begin{equation}
\overline{t}=\frac{e^{-ik_{\alpha}b}}{{\cos{k'_{\alpha}b}-i\mu \sin{k'_{\alpha}b}}}
\label{tl_forbidden_sSFQM}
\end{equation}
where,
\begin{eqnarray}
k'_{\alpha} &=&(-1)^{\frac{1}{\alpha}} \left( \frac{V-E}{D_{\alpha}}\right)^{\frac{1}{\alpha}} \nonumber  \\ 
			&=&q_{\alpha}e^{i\frac{\pi}{\alpha}}
\end{eqnarray}
and,
\begin{equation}
q_{\alpha}=\left( \frac{V-E}{D_{\alpha}}\right)^{\frac{1}{\alpha}}
\label{q_alpha}
\end{equation}
In order to find the tunneling time one needs to obtain the phase of the tunneling coefficient given by Eq. \ref{tl_forbidden_sSFQM}. The algebra for this is illustrated below.
\paragraph{}
In terms of $q_{\alpha}$ we can separate real and imaginary parts of $\sin{k'_{\alpha}b}$ and $\cos{k'_{\alpha}b}$.
\begin{equation}
\sin{k'_{\alpha}b}=\sin{\left( q_{\alpha}b \cos{\frac{\pi}{\alpha}}\right)}\cosh{\left( q_{\alpha}b \sin{\frac{\pi}{\alpha}}\right)}+i\cos{\left( q_{\alpha}b \cos{\frac{\pi}{\alpha}}\right)}\sinh{\left( q_{\alpha}b \sin{\frac{\pi}{\alpha}}\right)}
\label{sinkd_b}
\end{equation} 
\begin{equation}
\cos{k'_{\alpha}b}=\cos{\left( q_{\alpha}b \cos{\frac{\pi}{\alpha}}\right)}\cosh{\left( q_{\alpha}b \sin{\frac{\pi}{\alpha}}\right)}-i\sin{\left( q_{\alpha}b \cos{\frac{\pi}{\alpha}}\right)}\sinh{\left( q_{\alpha}b \sin{\frac{\pi}{\alpha}}\right)}
\label{coskd_b}
\end{equation} 
The real and imaginary parts of $\mu$ is written as
\begin{equation}
\mu=\frac{1}{2} \Big[\left( \varepsilon_{\alpha}+\frac{1}{\varepsilon_{\alpha}}\right)\cos{\frac{\alpha-1}{\alpha}} \pi  -i \left( \varepsilon_{\alpha}-\frac{1}{\varepsilon_{\alpha}}\right)\sin{\frac{\alpha-1}{\alpha}} \pi  \Big]
\label{mu_real_img}
\end{equation}
where,
\begin{equation}
\varepsilon_{\alpha}=\left(  \frac{k_{\alpha}}{q_{\alpha}} \right)^{\alpha-1}
\end{equation}
With the help of Eqs. \ref{sinkd_b}, \ref{coskd_b} and \ref{mu_real_img} we  simplify
\begin{equation}
\cos{k'_{\alpha}b}-i\mu \sin{k'_{\alpha}b}=X-iY
\end{equation}  
where $X$ and $Y$ are
\begin{multline}
X=\cos{\big( q_{\alpha}b \cos{\frac{\pi}{\alpha}}\big)}\cosh{\big( q_{\alpha}b \sin{\frac{\pi}{\alpha}}\big)} \\
-\frac{1}{2}\Big[ \Big(\varepsilon_{\alpha}+\frac{1}{\varepsilon_{\alpha}} \Big) \cos{\big( q_{\alpha}b \cos{\frac{\pi}{\alpha}}\big)}\sinh{\big( q_{\alpha}b \sin{\frac{\pi}{\alpha}}\big)} \cos{\frac{\alpha-1}{\alpha} \pi}  \\ 
-\Big(\varepsilon_{\alpha}-\frac{1}{\varepsilon_{\alpha}} \Big) \sin{\big( q_{\alpha}b \cos{\frac{\pi}{\alpha}}\big)}\cosh{\big( q_{\alpha}b \sin{\frac{\pi}{\alpha}}\big)} \sin{\frac{\alpha-1}{\alpha} \pi}
 \Big]
 \label{X_expression}
\end{multline}
\begin{multline}
Y=-\sin{\big( q_{\alpha}b \cos{\frac{\pi}{\alpha}}\big)}\sinh{\big( q_{\alpha}b \sin{\frac{\pi}{\alpha}}\big)} \\
+\frac{1}{2}\Big[ \Big(\varepsilon_{\alpha}+\frac{1}{\varepsilon_{\alpha}} \Big) \sin{\big( q_{\alpha}b \cos{\frac{\pi}{\alpha}}\big)}\cosh{\big( q_{\alpha}b \sin{\frac{\pi}{\alpha}}\big)} \cos{\frac{\alpha-1}{\alpha} \pi}  \\ 
+\Big(\varepsilon_{\alpha}-\frac{1}{\varepsilon_{\alpha}} \Big) \cos{\big( q_{\alpha}b \cos{\frac{\pi}{\alpha}}\big)}\sinh{\big( q_{\alpha}b \sin{\frac{\pi}{\alpha}}\big)} \sin{\frac{\alpha-1}{\alpha} \pi}
 \Big]
 \label{Y_expression}
\end{multline}
With the help of Eqs. \ref{X_expression} and \ref{Y_expression} the net tunneling phase $\Phi(E)$ of Eq. \ref{tl_forbidden_sSFQM} is obtained as
\begin{equation}
\Phi=\theta-k_{\alpha}b
\end{equation}
where $\theta$ is
\begin{equation}
\theta=\tan^{-1}{\Big( \frac{Y}{X} \Big)}
\end{equation}
To find the phase delay time we evaluate $\frac{d\theta}{dE}$ now. 
\begin{equation}
\frac{d\theta}{dE}=\frac{d_{\alpha}}{(\cos^{2}{k'_{\alpha}b}+\mu^{2}\sin^{2}{k'_{\alpha}b})}
\label{dtheta_dE}
\end{equation}
\begin{equation}
d_{\alpha}=\left( X\frac{dY}{dE}-Y\frac{dX}{dE}\right)
\end{equation}
The calculation for $d_{\alpha}$ is very lengthy  hence we only write the final expression.
\begin{multline}
d_{\alpha}= \frac{1}{2}b \varepsilon_{+}q'_{\alpha}\cos{\beta}\cos{\gamma}\cosh{2\xi}+\frac{1}{2}b \varepsilon_{-}q'_{\alpha}\sin{\beta}\sin{\gamma}\cos{2\eta}\\ 
+\frac{1}{4}\varepsilon'_{+}\cos{\beta}\sin{2\eta}-\frac{1}{2}b q'_{\alpha}(\sin{\gamma}\sin{2\eta}+\cos{\gamma}\sinh{2\xi})\\
+\frac{1}{8}b q'_{\alpha}(\sin{2\eta}\sin{\gamma}-\sinh{2\xi}\cos{\gamma})(\varepsilon_{+}^{2}\cos^{2}{\beta}+ \varepsilon_{-}^{2}\sin^{2}{\beta})\\
+\frac{1}{8}\sin{2\beta}(\cosh^{2}{2\xi}\sin^{2}{\eta}+\sinh^{2}{2\xi}\cos^{2}{\eta})(\varepsilon_{-}\varepsilon'_{+}-\varepsilon_{+}\varepsilon'_{-})\\
+\frac{1}{4}\varepsilon'_{-}\sin{\beta}\sinh{2\xi}
\label{xdy_ydx}
\end{multline} 
The denominator of Eq. \ref{dtheta_dE} is 
\begin{multline}
v_{\alpha}=\cos^{2}{k'_{\alpha}b}+\mu^{2}\sin^{2}{k'_{\alpha}b}= \\
\frac{1}{16}\Big[ \{ 8-\varepsilon_{-}^{2}-\varepsilon_{+}^{2} -(\varepsilon_{+}^{2}- \varepsilon_{-}^{2})\cos{2\beta} \}\cos{2\eta} + \\
\{ 8+\varepsilon_{-}^{2}+\varepsilon_{+}^{2} +(\varepsilon_{+}^{2}- \varepsilon_{-}^{2})\cos{2\beta}\} \cosh{2\xi}   \\
+8\varepsilon_{-}\sin{\beta}\sin{2\eta}- 8\varepsilon_{+}\cos{\beta}\sinh{2\xi} \Big]
\label{valpha}
\end{multline} 
Here,
\begin{equation}
\eta=q_{\alpha}b\cos{\frac{\pi}{\alpha}} , \quad   \xi=q_{\alpha}b\sin{\frac{\pi}{\alpha}} , \quad   \beta=\frac{\alpha-1}{\alpha}\pi, \quad  \gamma=\frac{\pi}{\alpha}
\end{equation}
and,
\begin{equation}
q'_{\alpha}=\frac{dq_{\alpha}}{dE}=-\frac{1}{\alpha D_{\alpha}}q_{\alpha}^{1-\alpha}
\label{q_dash}
\end{equation}
\begin{equation}
\varepsilon_{\pm}=\varepsilon_{\alpha}\pm\frac{1}{\varepsilon_{\alpha}}
\label{eps_plus_minus}
\end{equation}
\begin{equation}
\varepsilon'_{\pm}=\frac{d \varepsilon_{+} }{dE}=\frac{\alpha-1}{\alpha} \frac{V}{(V-E)^{2}} \varepsilon_{\alpha}^{\frac{1}{1-\alpha}} (1\mp\varepsilon_{\alpha}^{-2})
\label{eps_plus_dash}
\end{equation}
Also,
\begin{equation}
\frac{dk_{\alpha}}{dE}=\frac{k_{\alpha}^{1-\alpha}}{\alpha D_{\alpha}}
\label{dkalpha_dk}
\end{equation}
Therefore the tunneling time from a rectangular barrier in space fractional quantum mechanics is given by
\begin{equation}
\tau_{\alpha}=\hbar \left(\frac{d_{\alpha}}{v_{\alpha}}-\frac{b k_{\alpha}^{1-\alpha}}{\alpha D_{\alpha}} \right)+\frac{b}{\frac{\hbar k}{m}}
\label{tau_alpha}
\end{equation} 
Where $d_{\alpha}$ and $v_{\alpha}$ are obtained from Eqs. \ref{xdy_ydx} and \ref{valpha} respectively.
\paragraph{}
Next we check the behavior of tunneling time in the limit $b \rightarrow \infty$ to see the possibility of existence of Hartman effect. By using the fact
\begin{equation}
\lim_{\rho \to \infty} \sinh{\rho}\sim \frac{1}{2}e^{\rho} \sim \lim_{\rho \to \infty} \cosh{\rho}
\end{equation}
we simplify
\begin{multline}
\lim_{b \to \infty} d_{\alpha} \sim \frac{1}{2}e^{2\phi}\big[ b q'_{\alpha} \varepsilon_{+}\cos{\beta}\cos{\gamma} +  \frac{1}{4}\sin{2\beta}(\varepsilon_{-} \varepsilon'_{+}-\varepsilon'_{-} \varepsilon_{+}) - \\
\frac{1}{4} b q'_{\alpha}\cos{\gamma} e^{2\phi}(\varepsilon^{2}_{+}\cos{\beta}^{2}+\varepsilon^{2}_{-}\sin{\beta}^{2})+\frac{1}{2}\varepsilon'_{-}\sin{\beta} -
b q'_{\alpha} \cos{\gamma} \big]
\label{dalpha_inf}
\end{multline}
\begin{multline}
\lim_{b \to \infty} v_{\alpha} \sim \frac{1}{2}e^{2\phi}\big[ 1+ \frac{1}{8} \{ \varepsilon_{+}^{2} +\varepsilon_{-}^{2} +(\varepsilon_{+}^{2} -\varepsilon_{-}^{2})\cos{2\beta} \} -\varepsilon_{+}\cos{\beta} \big]
\label{valpha_inf}
\end{multline}
From Eqs. \ref{dalpha_inf} and \ref{valpha_inf} it is clear that the exponential dependent part of width `$b$' cancels out in the expression $\lim_{b \to \infty} \frac{d_{\alpha}}{v_{\alpha}}$. This expression can be written in two parts: $b$-dependent part and $b$-independent part as follows:
\begin{equation}
\lim_{b \to \infty}  \frac{d_{\alpha}}{v_{\alpha}} \sim b q'_{\alpha}\tau_{1\alpha}+ \tau_{2\alpha}
\label{limiting_b}
\end{equation}
where $\tau_{1\alpha}$ and $\tau_{2\alpha}$ are given by
\begin{equation}
\tau_{1\alpha}=\frac{\cos{\gamma}\big[(\varepsilon_{+} \cos{\beta}-1)-\frac{1}{4} (\varepsilon_{+}^{2}\cos{\beta}^{2}+\varepsilon_{-}^{2}\sin{\beta}^{2} )\big]  }{  1+ \frac{1}{8} [ \varepsilon_{+}^{2} +\varepsilon_{-}^{2} +(\varepsilon_{+}^{2} -\varepsilon_{-}^{2})\cos{2\beta} ] -\varepsilon_{+}\cos{\beta}    }
\label{tau1_alpha}
\end{equation}    
\begin{equation}
\tau_{2\alpha}=\frac{\frac{1}{4}\sin{2\beta} (\varepsilon_{-}\varepsilon'_{+}-\varepsilon'_{-}\varepsilon_{+})+\frac{1}{2}\varepsilon'_{-}\sin{\beta}  }{  1+ \frac{1}{8} [ \varepsilon_{+}^{2} +\varepsilon_{-}^{2} +(\varepsilon_{+}^{2} -\varepsilon_{-}^{2})\cos{2\beta} ] -\varepsilon_{+}\cos{\beta}    }
\label{tau2_alpha}
\end{equation}  
Therefore for large $b$, the tunneling time expression (Eq. \ref{tau_alpha}) becomes
\begin{equation}
\lim_{b \to \infty} \tau_{\alpha} \sim b\left( q'_{\alpha} \tau_{1\alpha}- \frac {k_{\alpha}^{1-\alpha}}{\alpha D_{\alpha}}+\frac{1}{2k} \right) +\tau_{2 \alpha}
\label{tau_b_inf}
\end{equation}
This linearly depends on the  width of the barrier $b$. Therefore the Hartman effect doesn't exist in space-fractional quantum mechanics.  This is demonstrated in Fig \ref{fig_alpha} where  the variation of tunneling time  with the width of the barrier for various values of $\alpha$ is recorded. From the Fig \ref{fig_alpha} it is evident that as $b$ increases tunneling time increases to a maximum value $\tau_{max}$ at some value of the width $b_{max}$ (say) for different values of $\alpha$. The peak tunneling time can be obtained by maximizing $\tau$ with respect to $b$. This involves transcendental equation and therefore has to be done numerically. The variation of $\tau_{max}(\alpha, b_{max})$ with $\alpha$ along with $b_{max}$ is shown in Fig \ref{fig_peak}. Other parameters are same as of Fig \ref{fig_alpha}. It is observed that as $\alpha$ decreases, both $\tau_{max}$ and $b_{max}$ reduces.  

\subsection{Restoration of Hartman effect in standard quantum mechanics}
The known results of standard quantum mechanics can be recovered with $\alpha=2$. For $\alpha=2$ we have $\beta=\frac{\pi}{2}$ and $\gamma=\frac{\pi}{2}$. With these values, $\tau_{1 \alpha}=0$. We choose the unit $\hbar=1$, $c=1$ and  $2m=1$. Thus
\begin{equation}
k=k_{2}=\sqrt{E} \nonumber
\end{equation}
\begin{equation}
q=q_{2}=\sqrt{E-V} \nonumber
\end{equation}
We take the generalized diffusion coefficient $D_{\alpha}$ as \cite{tare_2014} 
\begin{equation}
D_{\alpha}  =\frac{u^{2-\alpha}}{\alpha m^{\alpha-1}}
\label{dalpha}
\end{equation}
where $u$ is the characteristic velocity of the non-relativistic system. Therefore in the chosen unit
\begin{equation}
D_{2}=1  \nonumber
\end{equation}
With the above substitution the tunneling time in standard quantum mechanics from a  rectangular barrier of large width reduces to 
\begin{equation}
\lim_{b \to \infty} \tau_{2} \sim \tau_{2\alpha} (\alpha=2)
\end{equation}
$\tau_{2\alpha}$ is independent of $b$. Therefore we obtain the Hartman effect of standard quantum mechanics from eq. \ref{tau_b_inf}. It can be worked out that
\begin{eqnarray}
\tau_{2\alpha}(\alpha=2) &= &\frac{\frac{1}{2}\varepsilon'_{-}  }{  1+ \frac{1}{8} [ \varepsilon_{+}^{2} +\varepsilon_{-}^{2} -(\varepsilon_{+}^{2} -\varepsilon_{-}^{2}) ]     }   \nonumber \\
 &= & \frac{1}{qk}
\end{eqnarray}
This is the exact result of standard quantum mechanics (See Eq. \ref{tt_qm}). \\
\section{Results and Discussions}
\label{results_discussions}
The stationary phase  method of calculating the tunneling time shows that the Hartman effect doesn't exist in space fractional quantum mechanics. The Hartman effect is recovered for $\alpha=2$ case i.e. the standard quantum mechanics. Variation of tunneling time with width of the barrier is shown in Fig \ref{fig_alpha}. It is seen that for $\alpha=2$ we have the Hartman effect. For $\alpha <2$ we see that the tunneling time depends on width `b' of the barrier. At first it increases with $b$ then begins to decrease as $b$ increases. Further the peak tunneling time decreases as $\alpha$ decreases (Fig \ref{fig_peak}). The reason behind these  paradoxial results are discussed below:
\paragraph{}
In space fractional quantum mechanics the path integral is taken over Levy flight paths. The Levy flight paths have more probability for particles to travel farther points over single jump as compared to Brownian paths. Therefore in SFQM a particle will take lesser time to cross the classically forbidden region than it would for standard quantum mechanics. This is indeed the case and is evident from the Fig \ref{fig_alpha}. However the decrease in tunneling time with increasing width  needs further explanation and has not been attempted in the current work. We conclude that in space fractional quantum mechanics  the Hartman  effect doesn't exist. After a certain width of classically opaque barrier the tunneling time monotonically decreases with the thickness of the barrier in SFQM which needs further investigations. Further this would be interesting to see the variation of tunneling time in time fractional quantum mechanics. This is yet to be investigated how the tunneling time varies with the variation of time fractional index $\beta$ for a given $\alpha$ in time fractional quantum mechanics.\\
\\
{\it \bf{Acknowledgements}}: \\
MH is thankful to Dr. Anil Agarwal, GD, SAG and Dr. M. Annadurai, Director, ISAC for their support to carry out this research work. BPM acknowledges the support from CAS, Department of Physics, BHU.


\begin{thebibliography}{1}
\bibitem {feynman} R. P. Feynman and A. R. Hibbs, Quantum Mechanics and Path Integrals ( McGraw-Hill, New York) 1965. 
\bibitem {1} N. Laskin, Phys. Lett. A 268, 298 (2000).
\bibitem {2} N. Laskin, Phys. Rev. E 62, 3135 (2000).
\bibitem {3} N. Laskin, Phys. Rev. E 66, 56108 (2002).
\bibitem {4} X. Guo, M. Xu, J. Math. Phys. 47, 82104 (2006).
\bibitem {5} N. Laskin, Comm. Nonlinear Sc. and Num. Sim. 12, 2 (2007).
\bibitem {6} J. Dong, M. Xu, J. Math. Phys. 48, 72105, (2007). 
\bibitem {longhi_fractional_optics} S. Longhi, Optics Lett. 40, 1117 (2015)
\bibitem {condon} E.U. Condon, Rev. Mod. Phys. 3, 43 (1931).
\bibitem {eisenbund_dissertation} L.P. Eisenbud, {\it Dissertation, Princeton, (unpublished)} (1948).
\bibitem {wigner_1955}  E.P. Wigner, Phys. Rev. 98, 145 (1955).
\bibitem {david_bohm_1951} D. Bohm, Quantum Theory, Prentice-Hall, New York (1951).
\bibitem {hartman_paper} T. E. Hartman, J. App. Phys. 33, 3427 (1962).
\bibitem {fletcher} J. R. Fletcher, J. Phys. C, 18, L55 (1985).
\bibitem {nimtz} G. Nimtz H. Spieker, H.M. Brodowsky, Phys. Lett. A 222, 125 (1996).
\bibitem {ph} Ph. Balcou and L. Dutriaux Phys. Lett. A,78, 851 (1997).
\bibitem {sattari} F. Sattari and E. Faizabadi AIP Advances 2, 12123 (2012) .
\bibitem {longhi1} S. Longhi, M. Marano, P. Laporta, and M. Belmonte Phys. Rev. E, 64, 055602 (2001).
\bibitem {generalized_hartman} V. S. Olkhovsky1, E. Recami and G. Salesi, Euro. Phys. Lett. 57, 879 (2002) 
\bibitem {esposito_multi_barrier} S. Esposito,  Phy. Rev. E 67, 016609 (2003)
\bibitem {dutta} A. Paul, A. Saha, S. Bandopadhyay and B. Dutta-Roy, Euro. Phys. Jour. D, 42, 495 (2007).
\bibitem{our} Ananya Ghatak, Mohammad Hasan, Bhabani Prasad Mandal, {\it Hartman-Fletcher effect for array of complex barriers}, arXiv: 1505.03163.
\bibitem {8} E. C. de Oliveira, J. V. Jr, J. Phys. A: Math. Theor. 44 (2011) 185303.


\bibitem {griffith} D. J. Griffiths and C. A. Steinkea, Am. J. Phys. 69 (2) (2001), 137-154.
\bibitem {tare_2014} J. D. Tare, J.P.H. Esguerra , Physica A 407 (2014) 43-53.
\bibitem {dutta_roy_book} “{\it Elements of Quantum Mechanics}”, B. Dutta Roy, {\it New Age Science Ltd.} (2009).

\end{thebibliography}
\end{document}